\documentclass[11pt,twoside]{article}

\usepackage{asp2006}
\usepackage{epsf}
\usepackage{lscape}

\begin{document}
\title{Pathways through interstellar matter: \\
From the closest stars to the most distant quasars}   
\author{Juan Vladilo}   
\affil{Osservatorio Astronomico di Trieste - INAF - Italy}

\begin{abstract} 
%
%
%
%
%
Observations of quasar absorption systems  
relevant for studies of 
star formation at redshift $2 \la z \la 4$ are briefly reviewed.
Emphasis is given on the role played by dust 
in our understanding of the star formation history
of galaxies detected as absorption systems.  
Local interstellar studies are used as a reference for understanding the properties  
of high redshift interstellar media.  
\end{abstract}

\section{Introduction}

The interstellar medium is relevant for two galactic processes
discussed in the present conference:
(1) provides the gas that feeds {\em star formation}
and (2) collects the {\em feedback} from stellar radiation and ejecta.
The present contribution
is focussed  on interstellar observations  
obtained from absorption line spectroscopy
of nearby stars and distant quasars.
Thanks to their brightness,
nearby stars offer the possibility to study the local interstellar medium in detail
since they can be observed at high spectral resolution from space.
High resolution is
required because interstellar lines are generally narrow;
space instrumentation allows us
to detect the most important interstellar absorption lines, which lie 
in the ultraviolet spectral range. 
The local interstellar medium  serves as a paradigm
for interstellar studies of distant galaxies.
The diffuse gas of high redshift galaxies can be probed
by means of high resolution spectroscopy of quasars.
Thanks to the expansion of the Universe,
many ultraviolet interstellar absorption lines of 
the intervening galaxies
are redshifted  to the optical spectral range,
where they can be studied in detail in high resolution quasar spectra
collected with 10-m class telescopes.

The fact that this conference serves as a celebration of the astrophysical career
of John Beckman offers me the opportunity to recall,
in the first part of my talk, 
some of the local interstellar studies 
that I had the pleasure of doing in collaboration with John
during his first years  at the {\em Instituto de Astrof\'\i sica de Canarias}.
   In the second part I will review some recent work on the interstellar
media of high redshift galaxies observed as quasar absorption line systems.

\section{Pathways to nearby stars} 

Interstellar absorption lines were discovered accidentally, as the result
of studies of high resolution stellar spectroscopy:
\ion{Ca}{ii} interstellar lines were first identified   
in the spectrum of $\delta$ Ori 
because they did not share the photospheric radial velocity curve  
 of this binary star (Hartmann, 1904). 
It is less known that the \ion{Mg}{ii} interstellar lines of the very local gas
($\la 10$ pc from the Sun) were discovered accidentally 
in the \ion{Mg}{ii} chromospheric emissions of the 
cool star $\beta$ Hyi, in the framework of a collaboration
between John Beckman, newly arrived at the IAC,
and the Trieste group of stellar spectroscopy  (Vladilo et al. 1985). 
The claim for detection of interstellar \ion{Mg}{ii} was based
 on the velocity separation between
the  photospheric/chromospheric lines and the projected velocity vector 
of the interstellar medium, predicted from previous studies of the
 gas within $\simeq 10^2$ pc from the Sun (Crutcher 1982).  
Observations taken with the {\em International Ultraviolet Explorer} (IUE)
confirmed the interstellar nature
of the \ion{Mg}{ii} absorptions detected at the edges of the stellar chromospheric emissions
 (Molaro, Vladilo \& Beckman 1986). 
Having developed this method of identification of local \ion{Mg}{ii} lines
in late-type stars
\footnote
{
Early-type stars provide a better background for the detection of
interstellar absorptions  than late-type stars;
however, they are too rare within $\sim 30$ pc.  
},
we analysed the full IUE spectroscopic database
and produced a map
of the interstellar gas at $\la$ 30 pc from the Sun
(Genova et al. 1990). 
We found that the nearby gas is very inhomogeneous, 
with regions of low density, in the direction of the Galactic anticenter and
the Galactic poles, and regions of high density, in the direction of the Galactic center.

As a result of the  ISM kinematics and
of the inhomogeneous distribution of the gas, the local
\ion{Mg}{ii} insterstellar absorption shows remarkable spatial variations in 
radial velocity and intensity which confuse the analysis of the 
stellar \ion{Mg}{ii} chromospheric lines. 
As a by-product of our analysis of the local gas, we were able to perform a study
of the \ion{Mg}{ii} chromospheric emission profiles uncontaminated by interstellar
absorption (Vladilo et al. 1987).
A lesson     learnt from this investigation  is that 
one should beware of the subtle effects introduced by the  
intervening medium on the observed properties of any class of astronomical objects.
 If this is true for nearby stars,  one should be concerned about studies
of distant quasars,  as we 
discuss  in the next part of this talk. 

The velocity vector of the local interstellar medium    
is material which has been shocked and accelerated by stellar winds and supernovae 
associated with the Sco-Oph OB association (Crutcher 1982).
This is an example of how the interstellar medium bears the signature
of stellar feedback and therefore can be used to trace the history of star formation
in galaxies. In the next section we will show other examples of the connection
between physical properties of the interstellar medium and 
star formation.

\section{Pathways to distant quasars}   

As originally predicted by Bahcall \& Salpeter (1965),
quasar spectra show absorption line systems   originating in the diffuse  
  gas  that  lie along the path to the quasar. 
The   study of these spectra allows us to
probe the physical and chemical properties of the intergalactic/interstellar medium
over a large fraction of the Hubble time, back to the epoch of quasar formation,
with an accuracy comparable to that achievable in UV
studies of the local interstellar medium mentioned in the previous section. 

Quasar absorbers are classified according to their 
 \ion{H}{i} column density, $N$(\ion{H}{i}), and to the presence of metal lines at
the same redshift of the Ly\,$\alpha$, if any.
The weak  Ly\,$\alpha$ lines, with  column densities  
$10^{12}  \la N$(\ion{H}{i}) [atoms cm$^{-2}$] $\la 10^{16}$  
are very numerous and 
produce a ``forest'' of narrow absorptions in the spectral region
bluewards of the Ly\,$\alpha$ emission;
no metal lines are directly associated with these absorbers. 
At the other extreme of the column density distribution, we have
the strong Ly\,$\alpha$ lines with $N$(\ion{H}{i}) $\geq 2 \times 10^{20}$  atoms cm$^{-2}$,
which are instead very rare and 
always associated with metal lines of high and low ionization species. 
These absorbers are called ``damped Ly $\alpha$'' (DLA) systems 
(Wolfe et al. 1986), because 
the Ly $\alpha$ line profile shows   ``damping wings'',
due to natural broadening, that extend beyond the doppler core.
There is common consensus that DLA systems originate in the interstellar medium
of intervening galaxies (hereafter ``DLA galaxies''). 
A recent review on DLA systems can be found in Wolfe, Gawiser \& Prochaska (2005).
Here we focus on  the effects of dust on our understanding of 
the star formation history of DLA galaxies.

By analogy with studies of nearby galaxies, we expect interstellar dust
to be a pervasive component of DLA galaxies.
The observational evidence for the existence of
dust in DLA systems is slowly growing in the course of the years. 
At low redshift,  definitive evidence  
has been found in  the DLA system at $z=0.52$ towards AO 0235+164,
where the dust extinction bump at 2175\,\AA\
and the silicate  absorption at 9.7\,$\mu$ have been detected 
(Junkkarinen et al. 2004; Kulkarni et al. 2007). 
The evidence for dust in the bulk of the DLA population
is mainly based  on studies of elemental depletions:
by analogy with local interstellar studies we expect  
refractory elements, such as Fe or Cr, to be depleted from the gas phase, 
where they can be detected via absorption spectroscopy, to the dust component, 
where they escape detection with this technique;
we instead expect volatile elements, such as Zn or S, to be undepleted.
Evidence for differential depletion in DLA systems have been confirmed by many studies
(Pettini et al. 1997, 2000; Hou, Boissier \& Prantzos 2001; 
Prochaska \& Wolfe 2002; Vladilo 1998, 2004;  
Dessauges-Zavadsky et al. 2006).
The existence of general
trends between depletion, metallicity and H$_2$ molecular fraction
(Ledoux, Petitjean, \& Srianand 2003; Petitjean et al. 2006) indicate that
  the observed depletions are indeed due to dust, rather than to 
anomalous DLA abundance patterns. 

To  probe the properties of the high redshift dust in  DLA galaxies 
it is important to complement the studies of differential
depletion  with studies of quasar extinction.
From of our knowledge of local dust, we expect the extinction   
to become more  efficient with decreasing  wavelength 
and to produce a reddening of the quasar continuum. 
The measurement of quasar reddening due to {\em individual} DLA systems is 
quite challenging (Ellison, Hall \& Lira 2005); 
reddening detections have been obtained so far for a few
metal rich systems at  redshift $z \la 2$  (Vladilo et al. 2006).
The detection of
the {\em mean} quasar reddening due to \ion{Ca}{ii} systems
(Wild, Hewett \& Pettini 2006)   
and \ion{Mg}{ii}   systems
(York et al. 2006; Menard et al. 2007) at $z \la 1$
suggests that also for DLA systems should it be possible 
to obtain a statistical detection.
A claim of  detection 
based on the analysis of a sample of 13 DLA/QSOs (Pei, Fall \& Bechtold 1991)  
was not confirmed by the study of a large set
of Sloan Digital Sky Survey (SDSS) spectra of the 2nd data release     
(Murphy \& Liske 2004).
Analysis of the photometric and spectroscopic database of
the 5th SDSS data release    
indicates that the mean quasar reddening due to DLA systems at $2.2 \la z \la 3.5$
has been detected; the low value of dust-to-gas ratio obtained,
$\langle A_V/N(\mathrm{HI}) \rangle \sim 3 \times 10^{-23}$ mag cm$^{-2}$,
is  in line with the low level of metallicity of these systems  
(Vladilo, Prochaska \& Wolfe 2007).

\subsection{Dust and star formation in DLA systems}

 The techniques used to determine the star formation rate
(SFR)  based on the measurement of emission lines
are   ineffective in DLA systems.
This is due to the difficulty of disentangling
the DLA galaxy against the quasar PSF.  
Only in a few DLA systems has it been possible to derive 
limits, 
 in the order
of a few  $M_{\odot}$ year$^{-1}$, from studies of the Ly\,$\alpha$  
(M\o ller, Fynbo \& Fall, 2004) or  H\,$\alpha$ (Kulkarni et al. 2001) emission.
A value larger by one order of magnitude
was derived in one DLA system at $z=1.9$  
from the analysis of the FUV continuum (M\o ller et al. 2002).
The paucity of these measurements indicates   the need of  
especially designed  techniques aimed at determining  the
SFR in DLA systems. 

A method developed by  Wolfe, Prochaska \& Gawiser (2003) makes use
of the column density of the \ion{C}{ii}$^\ast$ 1335.7\,\AA\ line 
to determine the [\ion{C}{ii}] 158 $\mu$m cooling rate.
Assuming thermal balance one can estimate the heating rate 
and so  constrain the efficiency of the heating mechanisms.
The main heating mechanism considered is photoelectric emission from dust grains.
The efficiency of this mechanism depends on the intensity of the interstellar
radiation field, which in turn is related to the SFR. 
With this method two SFR solutions can be found, corresponding to the thermally stable
states of a two-phase medium with a cold neutral medium (CNM) in pressure
equilibrium with a warm neutral medium (WNM).
The average SFR per unit area found for 23 DLA systems 
with detected \ion{C}{ii}$^\ast$  line 
is $\langle \dot{\psi}_\ast \rangle \simeq 11 \times 10^{-3}$ $M_{\odot}$ year$^{-1}$ kpc$^{-2}$
for the CNM solution; the larger value found for the WNM solution,
$\langle \dot{\psi}_\ast \rangle \simeq 0.21$ $M_{\odot}$ year$^{-1}$ kpc$^{-2}$,
implies a bolometric background in excess of the observational limits 
(Wolfe et al. 2004). 
Since the efficiency of the heating mechanism considered in these computations depends 
on the amount of dust present in the medium,
an accurate estimate of the  dust-to-gas ratio is necessary
to infer an accurate value of SFR from the \ion{C}{ii}$^\ast$ 1335.7\,\AA\ column densities. 
 
If the  Kennicutt-Schmidt law is valid at high redshift, one can estimate
the SFR in DLA systems from the measured  \ion{H}{i} column densities. 
This can be done by taking into account inclination effects to convert the  
$N$(\ion{H}{i}) into the column density perpendicular to the galactic disk, $N_\bot$,
which is used in the law $(\dot{\psi}_\ast)_\bot = K \, (N_\bot/N_c)^\beta$ (Kennicutt 1998).
From this type of calculation, Wolfe \& Chen (2006) find that 3\%  
of the sky should be covered with extended objects brighter than  $\mu_V  \sim 28.4$ mag arcsec$^{-2}$,
if DLAs at redshift $z \in [2.5,3.5]$  undergo in situ star formation.
To test this hypothesis these authors searched for low-surface brightness features in
the Hubble Ultra Deep Field (UDF).
They found upper limits on the comoving SFR densities 
that are between factors of 30 and 100 lower than predictions, 
suggesting a reduction by more than a factor
of 10 in the star formation efficiency predicted by the Kennicutt-Schmidt law
at $z \sim  3$.
This type of conclusion maybe affected  to some extent 
by the presence of dust:
 the extinction would yield a lower observed intensity and hence
a lower value of the constant $K$ inferred with this method.

Constraints on the SFR history of DLA galaxies can also be obtained
by comparing the elemental abundances measured from the spectra
with those predicted by the models
of galactic chemical evolution (e.g. Matteucci, Molaro \& Vladilo 1997). 
Abundance ratios can be used as diagnostics of evolution if they
involve two elements synthesized on different time-scales, such
as  $\alpha$-capture elements,
mainly produced by type II SNe, and iron-peak elements, mainly produced by type Ia SNe. 
The $\alpha$/Fe ratio ratio, when compared with a metallicity tracer, such as  Fe/H,
allow us to clarify the particular history of star formation involved. 
In a regime of high SFR we expect to observe enhanced   $\alpha$/Fe ratios
for a large interval of Fe/H, while the opposite is expected for a regime
of low SFR (Matteucci 1991). 
In applying this type of study to DLA systems we face the problem
that the measured abundance ratios may be affected by differential depletion.
In fact, the $\alpha$/Fe ratios measured in
DLA systems are often enhanced due to differential depletion,
as it is the case of the Si/Fe ratios.
This apparent enhancement mimics the behaviour of galaxies with relatively high SFR.
However, when depletion effects are taken into account (Vladilo 2002)
or volatile elements are used in the analysis (Centuri\'on et al. 2000),
or dust-free DLA systems are investigated (e.g. Molaro et al. 2000)
the resulting $\alpha$/Fe ratio is generally not enhanced. 
The comparison with chemical evolution models suggests an origin of DLA systems
in galactic regions with low or episodic SFR 
(Calura, Matteucci \& Vladilo 2003; Dessauges-Zavadsky et al. 2006).

The presence of dust may also affect our understanding of
the star formation history of DLA systems  
due to the extinction of the quasar continuum.  
In the most extreme cases,  the
extinction due to dust-rich DLA systems
could lead to the obscuration of the quasar  
(Ostriker \& Heisler 1984; Fall \& Pei 1989, 1993).
In  general, the extinction may induce a 
selection bias acting against the detection of dust-rich
galactic regions in magnitude-limited
surveys     (Boiss\'e et al. 1998; Prantzos \& Boissier 2000; Vladilo \& P\`eroux 2005).
Studies of radio-selected quasars surveys, not affected by extinction, 
suggest that the impact  of this effect 
on the statistical properties of DLA systems is modest
(Ellison et al. 2001; Jorgenson et al. 2006), 
but the size of these surveys is not large enough to reach
firm conclusions.
Due to the extinction bias, the mean DLA metallicity of 
magnitude-limited surveys   may be underestimated
since the dust-rich regions that are missed are likely to originate   
in  metal-rich systems.
In particular, the \ion{H}{i}-column density weighted metallicity 
$\langle \mathrm{M/H} \rangle_{w} =
[\sum_i ( \mathrm{M} / \mathrm{H})_i \times N_i(\mathrm{HI})]/ [\sum_i N_i(\mathrm{HI})]$,
an indicator of the mean cosmic metallicity at high redshift
(Lanzetta, Wolfe \& Turnshek 1995) may be underestimated.
The weighted metallicity $\langle \mathrm{M/H} \rangle_{w}$
can be used to infer the global rate of star formation of the Universe 
at high redshift  (Pei \& Fall 1995; Pei, Fall \& Houser 1999).
If $\langle \mathrm{M/H} \rangle_{w}$ is underestimated  
also the global SFR is underestimated. Present estimates of this bias
are still open to debate (Akermann et al. 2005, Vladilo \& P\'eroux 2005).

\begin{figure}[!ht]
 \plotfiddle{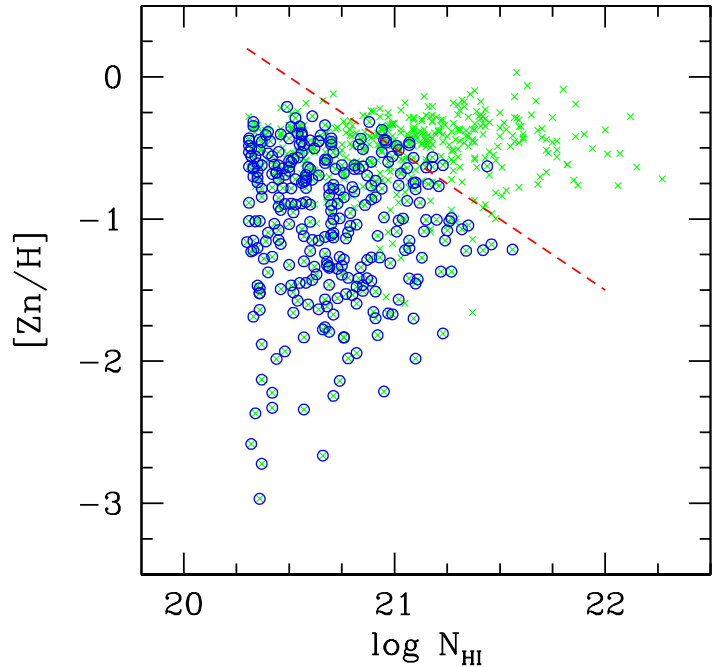}{9 cm}{0 }{50 }{50 }{-144 }{-90 }
\caption{Metallicity versus \ion{H}{i} column density in DLA galaxies predicted
by a cosmological simulation (Nagamine et al. 2004; model Q5).
Crosses:  systems extracted at random from the simulation.
Circles: systems observable in a DLA-QSO survey with limiting magnitude $r=19$.
An SMC-type extinction curve has been adopted to obtain the extinction in the
observer's frame.
Dashed line: empirical threshold [Zn/H] + log $N$(\ion{H}{i}) $=$ 20.5.
 }
\end{figure}

\subsection{Are we missing the regions of high SFR?}
  
The study of the metallicities versus \ion{H}{i} column densities 
indicates that DLA systems are concentrated {\em below} the value
of metal column density\footnote{
Zinc is used as a tracer of the metals since it is believed to be undepleted.
We adopt the usual definition of metallicity relative to the solar value
[X/H] $\equiv \log N(\mathrm{X})/N(\mathrm{H}) - \log (\mathrm{X/H})_\odot$. 
}
 [Zn/H] + log $N$(\ion{H}{i}) $\sim$ 20.5 (Boiss\'e et al. 1998).
Interstellar regions with
metal column densities {\em above} this empirical threshold
are instead predicted to exist at high redshift by models of galactic chemical
evolution (Prantzos \& Boissier 2000; Churches, Nelson \& Edmunds 2004)
and by SPH cosmological simulations of DLA galaxies (Cen et al. 2003; Nagamine et al. 2004). 
The lack of observed DLA systems at [Zn/H] + log $N$(\ion{H}{i}) $>$ 20.5 
has been attributed by some authors to the extinction bias, that would affect DLA systems
with high metallicity and $N$(\ion{H}{i})  
(Boiss\'e et al. 1998; Vladilo \& P\'eroux 2005). 
If this hypothesis is correct, we would be missing the regions of high star formation rate.
The relation between interstellar  extinction, $A_\lambda$ [mag], and metal column density  
is fundamental to test this hypothesis. 
Adopting iron as an indicator of metallicity, one can derive the following
expression for the $V$-band extinction 
in the rest frame of the absorber
\begin{equation}
A_V = \langle s_V^\mathrm{Fe}\rangle \times f_\mathrm{Fe} \times \mathrm{(Fe/H)} \times 
N(\mathrm{ HI}) ~,
\label{AVsV}
\end{equation} 
where 
(Fe/H) is the total abundance  by number of iron (gas plus dust),
$f_\mathrm{Fe}$ is the fraction of iron in dust form, and
$\langle s_V^\mathrm{Fe}\rangle$ is a line-of-sight average of dust grain parameters
(Vladilo et al. 2006). 
Empirical estimates of
 $\langle s_V^\mathrm{Fe}\rangle$  yield a typical
value $\langle s_V^\mathrm{Fe}\rangle \approx 3 \times 10^{-17}$ mag cm$^2$
in  \ion{H}{i} interstellar clouds,  metal absorption systems and a few
DLA systems   at $z_a \la 2$ (Vladilo et al. 2006). 
By adopting this value of  $\langle s_V^\mathrm{Fe}\rangle$,
together with an empirical estimate of  $f_\mathrm{Fe}$ (Vladilo 2004), 
we can use relation (\ref{AVsV}) to compute
the  extinction expected for DLA systems of given metal column density. 
This type of calculation can be applied, for instance, to test the 
predictions of SPH cosmological simulations. 
For each DLA system found at random in the cosmological box
one enters the
$N$(\ion{H}{i}) and (Fe/H) predicted by the simulation and
estimates the rest-frame extinction $A_V$.
The extinction in the observer's frame is then obtained adopting 
a normalized curve of interstellar extinction, $\xi(\lambda) = A_\lambda/A_V$.
This extinction is added to the magnitude of a simulated quasar,
which is assigned at random to each line of sight according to the
known frequency distribution of quasar magnitudes.
Lines of sight with dimmed magnitude in excess of a given 
magnitude limit are then discarded from the mock sample.
An example of result of this type of computation is shown in Fig. 1 
for a mock survey of DLA systems at $z \simeq 3$
with limiting magnitude $r=19$. This value is
representative of the current magnitude limit of  
high spectral resolution metallicity surveys.
One can see in Fig. 1 that the empirical threshold 
[Zn/H] + log $N$(\ion{H}{i}) $\la$ 20.5 
is naturally reproduced by the extinction effect.
This result  
suggests that DLA systems with high SFR are indeed   missed.
However, even after taking into account the extinction effect,
the mean weighted metallicity predicted
by the simulation, $ [ \langle \mathrm{M/H} \rangle_w] \simeq -0.85$ dex, 
is too high compared to the one measured in metallicity surveys
at $z \simeq 3$, $ [ \langle \mathrm{M/H} \rangle_w] \simeq -1.5$ dex
(Prochaska, Gawiser \& Wolfe 2001).
This discrepancy 
indicates that the SFR  predicted by the cosmological simulation is too high.
More refined simulations, with improved physics of star formation,
 are required  before this method can be used to 
quantify  the fraction of regions of high SFR missed due to the  extinction bias.

\acknowledgements 
It is a pleasure to thank the organizers, and in particular Johan Knapen,
for the invitation to be  part of a really eclectic experience.
I am grateful to John Beckman for the trust, encouragement and example
he gave me since the very beginning of my career.
This work has been supported by PRIN INAF 2006 (P.I. Molaro).

\end{document}